\documentclass[conference]{IEEEtran}
\usepackage{fancyhdr}
\IEEEoverridecommandlockouts

\usepackage{cite}
\usepackage[noend]{algorithm}
\usepackage[noend]{algpseudocode}
\usepackage{graphicx}
\usepackage{textcomp}
\usepackage{xcolor}
\usepackage[hyphens]{url}
\usepackage{amsmath, amssymb, amsthm}
\usepackage{bm}
\usepackage{float} 
\usepackage{subcaption}
\usepackage{booktabs}
\usepackage{multirow}
\usepackage{enumitem}
\usepackage{siunitx} 
\usepackage{hyperref}
\usepackage[normalem]{ulem}
\usepackage{tikz}

\newcommand*\circled[1]{\tikz[baseline=(char.base)]{
            \node[shape=circle,draw,inner sep=0.1pt] (char) {#1};}}

\def\BibTeX{{\rm B\kern-.05em{\sc i\kern-.025em b}\kern-.08em
    T\kern-.1667em\lower.7ex\hbox{E}\kern-.125emX}}
\begin{document}

\pdfpagewidth=8.5in
\pdfpageheight=11in

\newcommand{\iscasubmissionnumber}{127}
\newcommand{\sysname}{PIMCQG}

\pagenumbering{arabic}

\title{\LARGE \bf PIMCQG: Co-Designing Billion-Scale Graph-Based Approximate Nearest Neighbor Search for Processing-in-Memory}


\newcommand{\authorentry}[4]{%
\IEEEauthorblockN{#1}
\IEEEauthorblockA{#2\\#3\\#4}
}

\author{
\authorentry{Sitian Chen}{Hong Kong Baptist University}{Hong Kong, China}{csstchen@comp.hkbu.edu.hk}
\and
\authorentry{Yusen Li}{Nankai University}{Tianjin, China}{liyusen@nbjl.nankai.edu.cn}
\and
\authorentry{Yao Chen}{Huazhong University of Science and Technology}{Wuhan, China}{chenyao\_cs@hust.edu.cn}
\and
\authorentry{Minwen Deng}{Tencent}{Shenzhen, China}{danierdeng@tencent.com}
\and
\authorentry{Jintao Meng}{Shenzhen Institutes of Advanced Technology}{Shenzhen, China}{jt.meng@siat.ac.cn}
\and
\authorentry{Amelie Chi Zhou\textsuperscript{*}\thanks{\textsuperscript{*}Corresponding author.}}{Hong Kong Baptist University}{Hong Kong, China}{amelieczhou@comp.hkbu.edu.hk}
}



\maketitle
\thispagestyle{fancy} 
\lhead{} 
\rhead{} 
\chead{} 
\lfoot{\footnotesize{ SC26, November 15-20, 2026, Chicago, Illinois, USA \newline 979-8-3195-4789-7/26/\$31.00 \copyright 2026 IEEE}} 
\rfoot{} 
\cfoot{} 
\renewcommand{\headrulewidth}{0pt} 
\renewcommand{\footrulewidth}{0pt} 



\begin{abstract}

Graph-based Approximate Nearest Neighbor Search (ANNS) has emerged as the dominant approach for large-scale AI applications, offering superior accuracy-efficiency trade-offs over alternative methods. However, graph-based ANNS is inherently memory-bound due to frequent and irregular graph traversal. This constrains CPU throughput due to limited memory bandwidth, while GPUs lack sufficient HBM capacity for billion-scale indexes. Processing-in-Memory (PIM) offers massive internal bandwidth by co-locating computation with data, making it a natural fit for irregular, memory-bound workloads. However, directly mapping graph-based ANNS onto PIM exposes severe algorithm–architecture mismatches: restricted per-PU local memory, costly inter-PU communication, heavy coordination overhead, and weak compute capability, forcing existing PIM-based ANNS to rely on cluster-based algorithms that plateau at low recall.
We propose \sysname{}, an algorithm--hardware co-design framework through three optimizations:
1) a compact index reducing PIM-resident footprint by $14.5\times$,
2) an asynchronous pipelined scheduler fully utilizing host--PIM bandwidth, and
3) a multiplication-free distance kernel with $<$0.08\% recall loss.
Evaluations on three billion-scale datasets show up to $20\times$/$17.1\times$ throughput over CPU/GPU baselines, $129\times$ over prior PIM accelerators at high recall, and scale well across multi-node and emerging PIM architectures.

\end{abstract}

\section{Introduction}

Approximate Nearest Neighbor Search (ANNS) is a foundational primitive for large-scale AI services, including retrieval-augmented generation (RAG)~\cite{lewis2020retrieval,chen2022approximate,dong2024journey}, and recommendation systems~\cite{vairale2020recommendation, wang2023ems}. 
Given a high-dimensional query vector, ANNS retrieves the top-$k$ most similar vectors from a massive database while trading a small amount of accuracy for orders-of-magnitude efficiency gains. 
Among the major ANNS families (hash-based~\cite{zheng2016lazylsh, gong2020idec}, cluster-based~\cite{liu2024juno,jiang2023co,chen2024upanns,chen2024drim}, graph-based~\cite{gou2025symphonyqg, wang2024ndsearch,zhu2023processing,fu2017fast,chen2023finger,wang2025accelerating}), graph-based methods
have received extensive attention in both academia and industry due to their superior expected search \emph{accuracy} and \emph{efficiency} trade-offs.

\begin{figure}[t]
    \centering
    \includegraphics[width=1\linewidth]{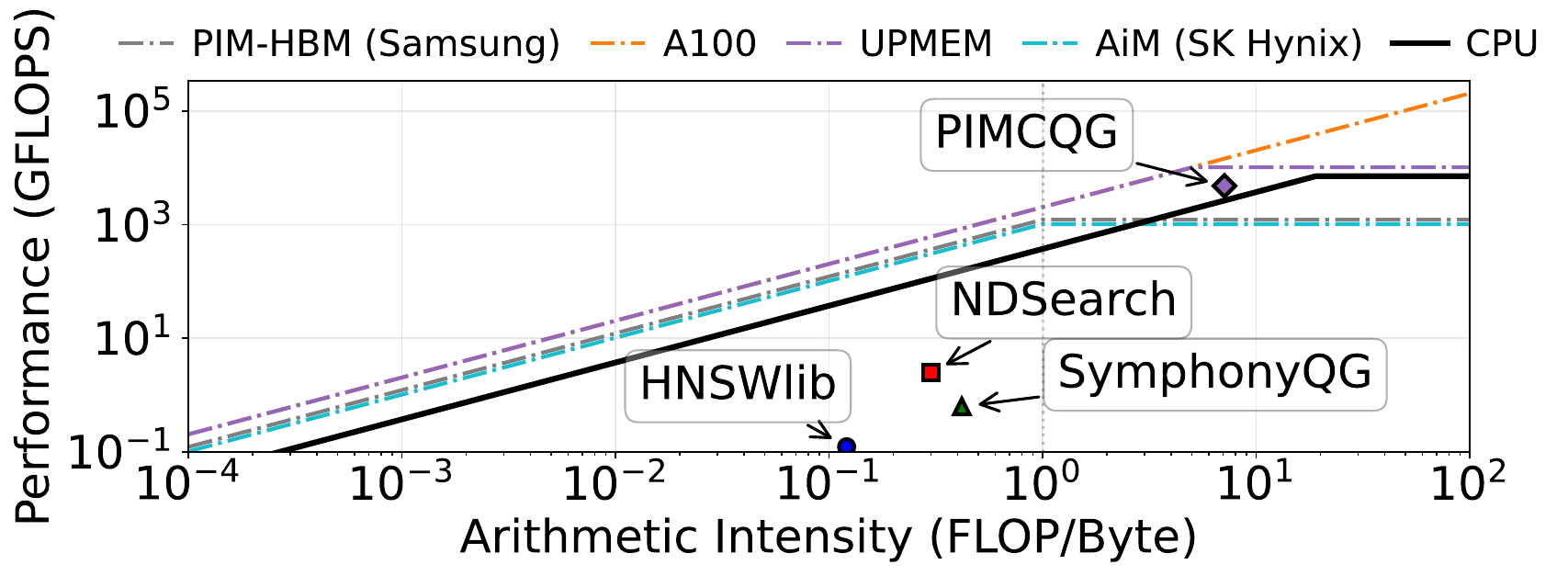}
    \caption{Roofline Analysis of graph-based ANNS on SIFT.}
    \label{fig:roofline}
\end{figure}

\textbf{Limitation of Existing Architectures.}
Despite algorithmic advances, graph-based ANNS on conventional architectures are fundamentally bottlenecked by the ``memory wall.'' 
To quantify this effect, we perform a roofline analysis of representative graph-based ANNS methods using the SIFT dataset and a CPU platform (2$\times$ Intel Xeon Gold 6330). As shown in Figure~\ref{fig:roofline}, the results indicate that these state-of-the-art methods, including HNSWlib~\cite{hnswlib}, NDSearch~\cite{wang2024ndsearch}, and SymphonyQG~\cite{gou2025symphonyqg}, lie firmly in the memory-bound region, struggling to effectively utilize available compute resources.
CPUs are constrained by limited memory bandwidth (e.g., 375~GB/s on CPUs), while GPUs, despite offering higher bandwidth (e.g., 2.0~TB/s on an A100), are severely limited by global memory capacity---a full SymphonyQG index for billion-scale datasets exceeds 1.25~TB, far beyond what GPU HBM can accommodate.

\textbf{Opportunities of PIM.}
To bridge this gap, Processing-in-Memory (PIM) has emerged as a promising architecture. 
Modern PIM systems (e.g., UPMEM PIM~\cite{gomez2022benchmarking}, Samsung PIM-HBM~\cite{kwon2022system}, and SK Hynix AiM~\cite{kwon202125}) embed lightweight processing units (PUs) directly inside memory banks. 
By co-locating computation with data, PIM enables massive internal memory bandwidth (e.g., 2~TB/s aggregate on UPMEM) and minimizes costly data movement across the memory hierarchy, making it particularly attractive for memory-bound, irregular workloads like graph traversal.
Unlike custom accelerator designs~\cite{zeng2023df, ko2025cosmos, song2025efficient, xu2026proxima}, commodity PIM modules ship as standard DIMM form-factor devices, enabling practical deployment without system redesign.
However, existing PIM-based ANNS accelerators~\cite{chen2024upanns,chen2024drim,wu2025turbocharge} predominantly rely on simpler, cluster-based algorithms (e.g., IVF-PQ), which inherently suffer from lower search quality and hit a capability ceiling at low recall levels (e.g., ${\sim}$61\% on SIFT1B~\cite{sift1b}).

\textbf{Challenges.}
While PIM's massive internal bandwidth makes it appear well suited for memory-bound workloads, directly mapping graph-based ANNS onto PIM exposes a fundamental algorithm--architecture mismatch that manifests as four tightly coupled challenges (detailed analysis in Section~\ref{sec:motivation}):
\circled{1} \textbf{Extreme Local Memory Capacity:} PU-private memory is severely limited (e.g., 64~MB per UPMEM DPU), yet billion-scale graph indexes exceed 1.25~TB, forcing aggressive partitioning.
\circled{2} \textbf{Inter-PU Communication:} Fine-grained partitioning causes frequent cross-PU traversals over external bandwidth paths that are $>$10$\times$ slower than internal bandwidth.
\circled{3} \textbf{Coordination Overhead and Load Imbalance:} Existing batch-synchronous~\cite{cui2025pimlex,cai2024pimpam,chen2024upanns,chen2024drim} and per-query~\cite{wu2025turbocharge} scheduling strategies either waste PU cycles on synchronization barriers or fragment communication bandwidth.
\circled{4} \textbf{Restricted PU Compute:} PIM cores feature weak arithmetic capabilities and in some cases lack hardware multipliers (e.g., UPMEM), resulting in disproportionate cost for the remaining floating-point operations in quantization-based distance kernels.
Overcoming these coupled challenges requires a holistic algorithm--hardware co-design.


\textbf{Innovations.}
To overcome these architectural constraints, we propose \sysname{}, a holistic algorithm--hardware co-design framework that enables high-performance, high-recall graph-based ANNS on commodity PIM platforms. 
Rather than applying isolated software patches, \sysname{} realigns the entire system stack---data layout, scheduling, and computation---with PIM architectural constraints through three synergistic innovations:
\circled{1} \textbf{PIM-Friendly Compact Index:} We eliminate redundant per-edge quantization metadata via IVF-style clustering and offload exact similarity reranking to the host CPU. This reduces the PIM-resident index footprint by up to $14.5\times$ (e.g., from 2,385~GB to 164~GB on SSN1B dataset~\cite{ssn1b}), dramatically easing capacity limits (\emph{Challenge 1}) and cross-PU communication pressure (\emph{Challenge 2}) (Section~\ref{sec:index}).
\circled{2} \textbf{Asynchronous Pipelined Scheduling:} We decouple host-side query dispatch and post-processing from in-PIM search through dynamic mini-batching and FIFO-based asynchronous execution. By overlapping communication, search, and reranking, this design mitigates coordination overhead and load imbalance (\emph{Challenge 3}) while fully utilizing host--PIM bandwidth (Section~\ref{sec:host-pu}).
\circled{3} \textbf{Multiplication-Free Distance Computation:} On observing that the quantization-error scaling factor $\cos(\theta)$ is empirically stable within each IVF cluster, we replace the expensive per-node floating-point multiplication with a cluster-wide constant approximated via bitwise shifts and additions, achieving a fully multiplication-free in-PIM distance kernel with $<$0.08\% recall loss (\emph{Challenge 4}) (Section~\ref{sec:multiply}).


\textbf{Evaluation.}
We evaluate \sysname{} on a real-world UPMEM PIM server across three industry-standard billion-scale datasets. 
Our results demonstrate three key advantages over state-of-the-art baselines:
1) \textbf{Performance:} \sysname{} delivers up to $20\times$ and $17.1\times$ throughput speedups over CPU-based SymphonyQG and GPU-based GGNN, respectively. Furthermore, unlike prior PIM-based ANNS solutions that hit a capability ceiling at low recall levels, \sysname{} successfully scales to high-recall targets, achieving up to $129\times$ speedup at comparable recall boundaries.
2) \textbf{Energy Efficiency:} \sysname{} provides up to $6.5\times$ and $30.8\times$ improvement in energy efficiency (QPS/Watt) compared to CPU and GPU baselines, demonstrating the power of PIM for large-scale ANNS.
3) \textbf{Scalability:} \sysname{} maintains its performance advantages when scaling to multi-node configurations and different commodity PIM platforms (Samsung PIM-HBM, SK Hynix AiM), showing its generality and practical deployment potential.
\section{Background and Motivation}
\subsection{Graph-based ANNS Algorithms} \label{sec:ganns}


\begin{figure}
\centering
        \includegraphics[width=.8\linewidth]{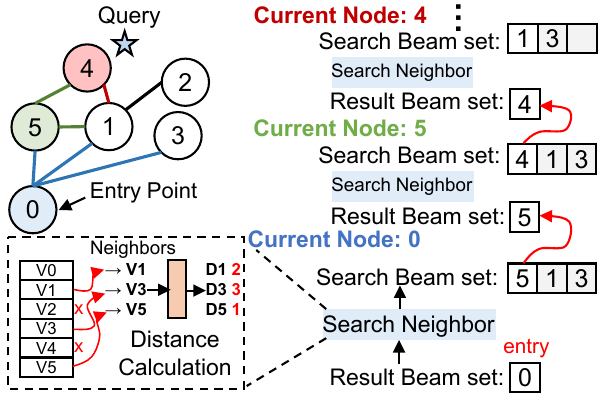}
    \caption{Query processing with greedy beam search.}\label{fig:beam_search}
\end{figure}

%
Graph-based ANNS approaches model the dataset as a proximity graph, enabling a greedy beam search~\cite{fu2021high, fu2017fast, jayaram2019diskann, malkov2018efficient} to navigate the structure. Starting from an entry point, 
the search maintains a candidate set (the \emph{beam}) of the current $n_b$ closest vectors and iteratively expands promising nodes by exploring their neighbors until convergence. Figure~\ref{fig:beam_search} illustrates this process for $n_b=3$ and $k$=1. The search begins at the entry node (node~0), fetches its neighbors, computes their distances to the query, and updates both the search and result beams accordingly.
Each iteration therefore consists of three tightly coupled actions:
\textbf{(1)} pointer-based graph traversal to access adjacency lists and candidate vectors,
\textbf{(2)} distance computation and ranking, and
\textbf{(3)} state updates to the beam and visited set.
This decomposition highlights a fundamental performance characteristic of graph-based ANNS: its efficiency depends on both irregular memory access from graph traversal and arithmetic cost from distance evaluation. 

Early designs were primarily bottlenecked by distance computation, motivating a large body of work on accelerating or approximating distance evaluation~\cite{YahooJapan, aguerrebere2023similarity, Zilliz, hnswlib}. 
Among them, RabitQ~\cite{gao2024rabitq} stands out as a state-of-the-art (SOTA) quantization technique, offering {strong theoretical error bounds} while maintaining high practical accuracy. It has been widely used in real-world systems and included in the FAISS library~\cite{faiss,shi2026gpu}.
Built upon this, SymphonyQG~\cite{gou2025symphonyqg} represents a SOTA graph-based ANNS design that tightly integrates RabitQ quantization with graph traversal.

SymphonyQG consists of two tightly coupled components, \textbf{offline index construction} and \textbf{online graph-based search}, both of which rely on RabitQ~\cite{gao2024rabitq} for quantization-aware distance estimation:
(1) The index construction stage builds a proximity graph over the dataset. For each node, it encodes its neighbors using RabitQ with respect to the current node, generating \emph{quantized codes} and associated scaling factors. This design enables efficient approximate distance computation during search, but also introduces additional metadata per edge (details in Figure~\ref{fig:index_structure}(a)).
(2) At query time, SymphonyQG performs greedy beam search over the graph as described in Figure~\ref{fig:beam_search}. Specifically, it computes {\uline{\emph{approximate} distances}} using the quantized representations generated by RabitQ, hence greatly reducing the computation cost. To maintain high recall, a small set of candidates (size of $n_b$) is periodically \uline{re-ranked using \emph{full-precision}} vectors. 
In this work, we focus on optimizing the query execution path, as it dominates end-to-end ANNS performance, while treating index construction as an offline preprocessing step. 


By aggressively reducing the cost of distance computation, modern graph-based ANNS algorithms \emph{{shift their performance bottleneck toward memory access and graph traversal}}, as confirmed by the roofline analysis in Figure~\ref{fig:roofline}.  
Motivated by this observation, this paper focuses on addressing the memory-bound nature of modern graph-based ANNS. We choose SymphonyQG as a representative algorithm to study and resolve the memory-related challenges.

\renewcommand{\arraystretch}{0.9}
\begin{table*}[t]
  \centering
  \caption{Comparison of representative CPU, GPU, and PIM platforms. The CPU, GPU, and UPMEM configurations are based on real hardware and are scaled to the same power budget for fair comparison. Specifications of PIM-HBM and AiM are derived from vendor technical reports, as no publicly available hardware is currently accessible.
           }
  \label{tab:hardware-comparison}
  
  \begin{tabular}{ l l l l l l l l }
    \toprule
    {Architecture} & {Type} & {Power (TDP)} & {Peak TFLOPS} & {External BW } & {Internal BW } & {\# PUs / Cores} & {PU-Private Memory} \\
    \midrule
    
    \multicolumn{8}{l}{\textbf{Power-Comparable Systems} (scaled to power budget of {$\sim$400}{W})} \\
    \addlinespace[0.3em] 
    
    UPMEM PIM     & PIM  & {$\sim$400}{W} & 14.0 (TOPS) & 150 GB/s & \textbf{2.8 TB/s} & 3584 DPUs    & 64 MB (MRAM) 
    \\
    (28 DIMMs) &      &  &            &      &              &              & \\
    
    Intel Xeon Gold 6330  & CPU  & {{410}{W}}   & 7.2      & 375 GB/s & {-}        & 56 Cores     & N/A (Shared DRAM) \\
    (Dual-Socket)    &      &  &            &      &              &              & \\
    
    NVIDIA A100      & GPU  & {{400}{W}}   & 312        & 2.0 TB/s  & {-}        & 6912 (CUDA)  & N/A (Shared HBM) \\
    (SXM4)           &      &                 &            &      &              & 432 (Tensor) & \\
    
    \addlinespace[0.4em] 
    
    \multicolumn{8}{l}{\textbf{Hypothetical / Component-Level Systems}} \\
    \addlinespace[0.3em]
    
    PIM-HBM (Samsung) & PIM  & {Not Stated}    & 1.2        & 307 GB/s & \textbf{1.2 TB/s} & 128 PUs      & 16 MB \\
    AiM (SK Hynix)     & PIM  & {Not Stated}    & 1.0        & 64 GB/s & \textbf{1.0 TB/s}  & 32 PUs       & 32 MB \\
    
    \bottomrule
  \end{tabular}
\end{table*}

\begin{figure}[t]
    \centering
    \includegraphics[width=.9\linewidth]{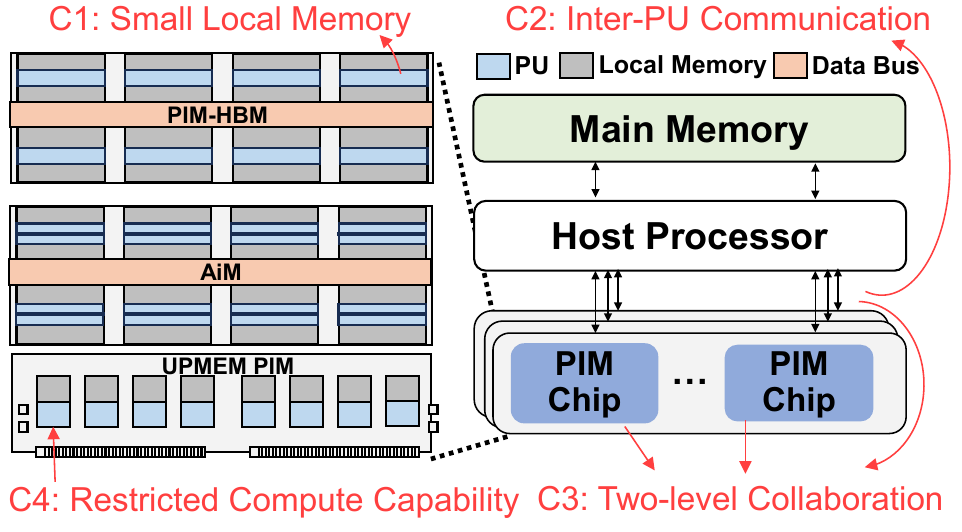}
    \caption{Commodity PIM architecture.}
    \label{fig:upmem}
\end{figure}

\subsection{Processing-In-Memory Architectures}\label{sec:upmem}


Processing-In-Memory (PIM) architectures represent a class of memory-centric systems designed to mitigate the long-standing memory wall by embedding lightweight processing units (PUs) directly inside or adjacent to memory banks. By co-locating computation with data, PIM enables high internal memory bandwidth and reduces costly data movement across the memory hierarchy.
Figure~\ref{fig:upmem} illustrates a representative commodity PIM architecture, capturing common design principles shared by modern PIM systems (e.g., UPMEM DRAM-PIM~\cite{gomez2022benchmarking}, Samsung PIM-HBM~\cite{kwon2022system} and SK Hynix AiM~\cite{kwon202125}).

A PIM system consists of a large number of memory chips, each augmented with multiple PUs. Every PU is tightly coupled to a private local memory bank and executes a lightweight instruction set, while a conventional host CPU remains responsible for query dispatch, synchronization, and result aggregation. 
Table~\ref{tab:hardware-comparison} further quantifies these architectural characteristics by comparing PIM systems with conventional CPU and GPU platforms. While PIM exposes \emph{significantly higher aggregate internal memory bandwidth} through massive parallelism across thousands of PUs, each PU operates with limited local memory capacity and relatively modest compute capability compared to modern processors.

\begin{figure*}[t]
    \centering
        \includegraphics[width=.95\linewidth]{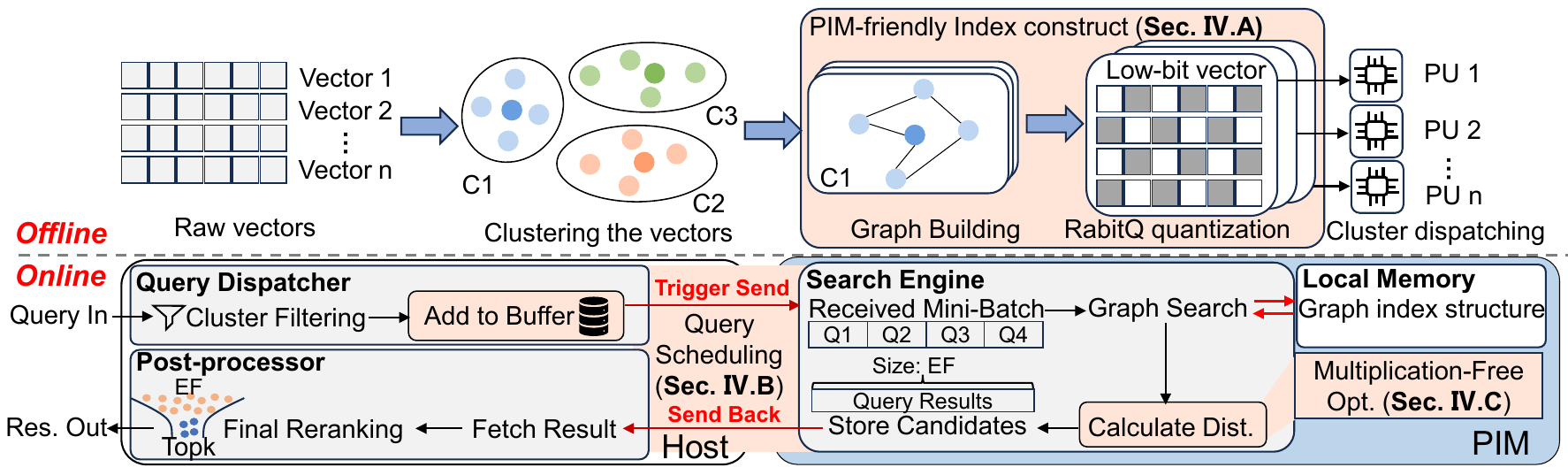}
        \caption{Overview of \sysname{}}
        \label{fig:overview}
\end{figure*}

\subsection{Co-Design Challenges for ANNS on PIM}\label{sec:motivation}

Although the massive internal bandwidth of PIM architectures makes them appear well suited for memory-bound workloads, applying PIM directly to graph-based ANNS exposes a fundamental algorithm–architecture mismatch. As discussed in Section~\ref{sec:ganns}, graph-based ANNS is inherently irregular and data-dependent, relying on dynamic graph traversal, evolving beams, and frequent state updates. At the same time, PIM architectures are distributed (data partitioned across thousands of PUs), resource-constrained (limited PU-private memory), and tightly coupled to host coordination. 
As a result, mapping modern graph-based ANNS onto PIM surfaces four tightly coupled challenges, which reflect the core reasons why existing PIM-based ANNS solutions have been limited to simpler, cluster-based algorithms~\cite{chen2024upanns,chen2024drim}.

\textbf{C1: Extreme Local Memory Capacity Constraints.}  
Each PU in a PIM system is equipped with only a small private memory bank (Table~\ref{tab:hardware-comparison}), yet modern graph-based ANNS indexes are memory intensive. Using SymphonyQG as an example, the index must store graph topology, quantization codes, and auxiliary metadata per node. For billion-scale datasets such as SIFT1B ($n=10^9$, $D=128$, graph degree $R=32$), the index footprint exceeds 1.25~TB, which would require partitioning the graph across tens of thousands of PUs. This extreme partitioning is not merely an engineering inconvenience; it fundamentally reshapes the execution behavior of graph traversal and directly exacerbates the next challenge.

\textbf{C2: Graph Partitioning vs. Inter-PU Communication.}  
The fine-grained partitioning forced by C1 splits the proximity graph across many PUs. During query processing, graph traversal frequently crosses partition boundaries. Any graph edge that crosses a boundary becomes a ``remote'' access that must traverse the slow external bandwidth path, which, as shown in Table~\ref{tab:hardware-comparison}, can be over an order of magnitude slower than the internal bandwidth. As a result, the benefits of PIM’s high internal bandwidth are nullified by communication overhead unless the index layout and traversal behavior are co-designed to minimize cross-PU interactions.


\textbf{C3: Coordination Overhead and Load Imbalance.} Communication overhead is further compounded by the highly data-dependent execution of graph-based ANNS. Because the proximity graph is statically partitioned across PUs, queries that traverse dense or frequently accessed partitions concentrate work on a small subset of PUs, while others remain underutilized. Existing PIM systems~\cite{cui2025pimlex, cai2024pimpam, li2024pim,chen2024upanns,chen2024drim} typically rely on rigid batch-synchronous execution models to amortize communication overheads, but these global barriers force fast PUs to idle while waiting for the slowest PU to complete a batch. Alternative fine-grained dispatching strategies (e.g., PIMANN~\cite{wu2025turbocharge}) reduce idle time but fragment communication, leaving aggregate bandwidth underutilized. Effectively addressing this challenge therefore requires rethinking how queries are scheduled and how host-side coordination is overlapped with in-PIM execution.

\textbf{C4: Restricted PU Compute Capability.}  
Even after memory, communication, and coordination issues are addressed, PIM processing units still provide orders of magnitude less compute throughput than GPUs, and in some designs lack hardware multipliers altogether (e.g. UPMEM). 
This constraint directly conflicts with the arithmetic patterns of graph-based ANNS, even in optimized methods like SymphonyQG, where approximate distance estimation still relies on multiplication-heavy inner products. Under such constraints, each arithmetic operation carries disproportionate cost, making it necessary to redesign the distance kernel.

These four challenges are deeply interrelated. Limited PU-local memory capacity (C1) forces aggressive graph partitioning, which in turn amplifies inter-PU communication during traversal (C2). High communication overhead exacerbates coordination costs and load imbalance across PUs (C3), while restricted PU compute capability (C4) further constrains the choice of distance computation and pruning strategies that could otherwise mitigate these effects. Overcoming them requires a holistic algorithm–hardware co-design. 

\section{\sysname{}: A CO-DESIGN FRAMEWORK FOR ANNS ON PIM}
\label{sec:overview}

This paper presents
\textbf{\emph{\sysname{}}}, a co-design framework that realigns state-of-the-art graph-based ANNS~\cite{gou2025symphonyqg} with the architectural realities of commodity PIM systems.
\sysname{} consists of three synergistic optimizations that jointly address the coupled challenges discussed earlier.

\textbf{O1: PIM-Friendly Compact Index.} 
To address \textbf{C1} and \textbf{C2}, we propose a compact index structure that removes redundant quantization metadata and offloads exact reranking to the host. 
This substantially reduces the memory footprint of the graph index, enabling massive datasets to fit within distributed PIM memories while also reducing the partitioning pressure that would otherwise amplify remote traversal and communication overhead. (Details in Section~\ref{sec:index}.)

\textbf{O2: Asynchronous Pipelined Scheduling.}
For \textbf{C3}, we propose an asynchronous pipelined scheduling architecture. It decouples host-side query dispatch and post-processing from in-PIM search through dynamic mini-batching and FIFO-based asynchronous execution. By overlapping communication, search, and post-processing, this design reduces synchronization stalls, better utilizes host--PIM bandwidth, and improves throughput under highly data-dependent workloads. (Details in Section~\ref{sec:host-pu}.)

\textbf{O3: Multiplication-Free Distance Computation.}
Even after memory and scheduling bottlenecks are mitigated,
\textbf{C4} remains a key obstacle. To address this, we redesign the distance computation kernel, and replace expensive, PIM-hostile floating-point multiplication operations with a sequence of highly efficient bitwise shift and addition operations. This better matches the strengths of lightweight PIM cores and enables efficient in-memory execution of graph-based ANNS without sacrificing search quality. 
(Details in Section~\ref{sec:multiply}.)

Figure~\ref{fig:overview} provides a high-level overview of \sysname{}’s end-to-end execution flow and illustrates how the proposed optimizations interact across the host and PIM. Conceptually, query processing is organized into three modules: a host-side \emph{Query Dispatcher}, an in-PIM \emph{Search Engine}, and a host-side \emph{Post-processor}. O1 defines the compact graph index traversed by the {Search Engine}, O2 governs the asynchronous interaction among modules, and O3 optimizes the distance computation used during candidate evaluation. The detailed design of each optimization is presented in Section~\ref{sec:design}.



\section{Design Details of \sysname{}}
\label{sec:design}


\subsection{PIM-friendly Compact Index}\label{sec:index}

\begin{figure}[t]
  \centering  
  \begin{subfigure}[b]{\linewidth}
    \centering
    \includegraphics[width=\linewidth]{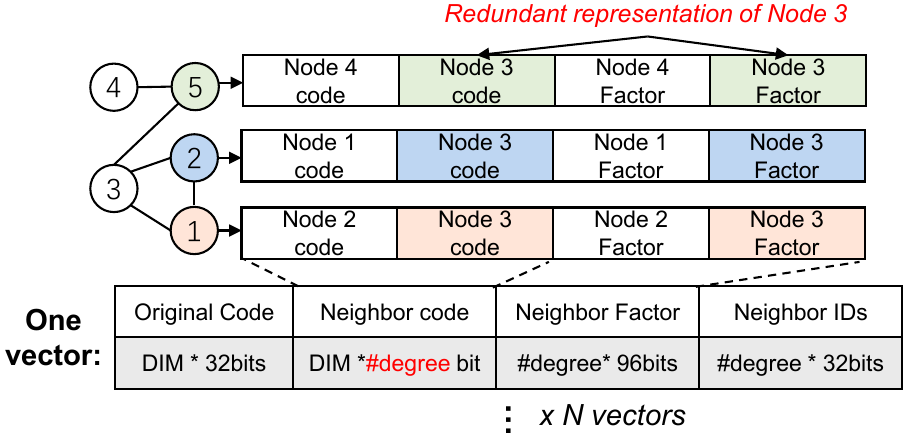}
    \caption{SymphonyQG (per-vector)}
  \end{subfigure}  
  \begin{subfigure}[b]{\linewidth}
    \centering
    \includegraphics[width=\linewidth]{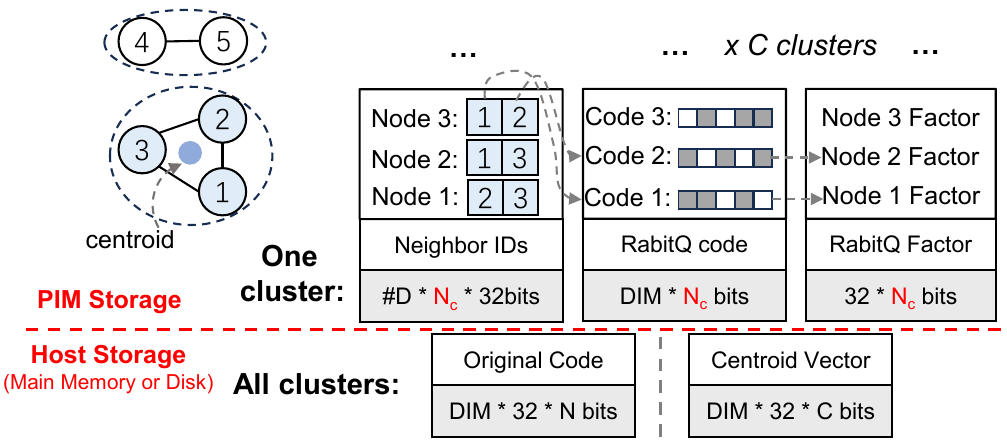}
    \caption{\sysname{} (per-cluster)}
  \end{subfigure}
  
  \caption{Index structure of SymphonyQG and \sysname{}}
  \label{fig:index_structure}
\end{figure}

We first revisit the internal organization of the SymphonyQG index.
As illustrated in Figure~\ref{fig:index_structure}(a), the index stores four parts per node, including the {original vector}, {neighbor codes}, {neighbor factors} and {neighbor IDs}. The ``neighbor code" and ``neighbor factor" components are the core of RabitQ quantization, enabling the replacement of expensive, full-precision distance calculations with highly efficient approximate ones. 
This design, while effective for enabling accurate quantization-aware distance estimation, introduces substantial redundancy and memory overhead. 
Therefore, directly porting the SymphonyQG index to PIM would not only exceed memory capacity, but also negate the benefits of near-data processing due to excessive communication overhead. 
To address these issues, we redesign the SymphonyQG index structure with two complementary techniques as described below.

\subsubsection{\textbf{Eliminating Index Redundancy}}


The redundancy is a direct result of SymphonyQG's quantization logic, where a neighbor's quantized code $\mathbf{o}$ is computed relative to the current node's vector $\mathbf{v}$ (i.e., $\mathbf{o} = \frac{\mathbf{o}_r - \mathbf{v}}{||\mathbf{o}_r - \mathbf{v}||}$), causing the same neighbor (e.g., ``Node 3" in Figure~\ref{fig:index_structure}(a)) to store different, redundant codes and factors for each incoming edge.

A naive approach to eliminating this redundancy would be to quantize all vectors using a single global reference. However, such a design fails to capture local data distributions and significantly degrades quantization accuracy. 
Instead, we adopt an {Inverted File (IVF)-style clustering} strategy: we partition the dataset into clusters, assign each node to a cluster, and use the cluster centroid $\mathbf{c}$ as the shared reference for quantization within the cluster (i.e., $\mathbf{o} = \frac{\mathbf{o}_r - \mathbf{c}}{||\mathbf{o}_r - \mathbf{c}||}$). Under this design, each node is encoded once relative to its assigned centroid, producing a \emph{single canonical} RabitQ code and scaling factor that can be reused by all incoming edges.

Figure~\ref{fig:index_structure}(b) shows the resulting PIM-aware index layout. The graph adjacency lists stored in PIM now contain only neighbor IDs, while the corresponding canonical RabitQ codes and scaling factors are stored once in shared PIM-resident arrays. During traversal, a PU follows a neighbor ID and resolves it to the node’s canonical code/factor entry, rather than edge-specific code/factor. Thus, multiple edges pointing to the same node reuse the same quantized representation, eliminating redundancy while preserving quantization accuracy.

\subsubsection{\textbf{Eliminating Raw Vectors via Approximate Ranking}}

After eliminating edge‑specific quantization redundancy, the remaining dominant contributor to index size in SymphonyQG is the storage of \emph{full‑precision raw vectors}. As described in Section~\ref{sec:ganns}, 
SymphonyQG retains full‑precision vectors to support exact distance computation during search. Although not used at every traversal step, these vectors are accessed periodically to rerank candidates and refine the beam, ensuring high recall.
However, storing raw vectors directly on PIM is impractical due to their large footprint ($\textit{DIM}*32$ bits per node). 

\sysname{} addresses this issue by restructuring the ranking pipeline to \emph{decouple approximate search from exact reranking}.
This design is motivated by two observations: 1) RabitQ‑based distance estimation already provides sufficiently accurate ordering for traversal and pruning; and 2) exact distances are only required for a small candidate set near convergence.
Therefore, \sysname{} removes raw vectors from the PIM‑resident index and retains them exclusively on the host. This enables \emph{approximate-only traversal on PIM}.
The query execution path is hence redesigned as follows: 
\begin{enumerate}[leftmargin=*]
    \item[1.] (\textbf{Host-side})  Filters the target clusters for the query and dispatches search requests to corresponding PUs.
    \item[2.] (\textbf{PIM-side})  Traverses the graph and computes approximate distances using canonical quantized representations.
    \item[3.] (\textbf{PIM-side})  Generates the candidate set with size $\textit{EF}$ based on approximate distances.
    \item[4.] (\textbf{Host-side}) Fetches the candidate set from PIM and performs a full-precision reranking to get the final top-$k$.
\end{enumerate}


\textbf{Accuracy tradeoff.}
Compared to the SymphonyQG baseline, \sysname{} quantizes vectors to their respective centroids, other than to the current searched node. This sacrifices locality and may introduce accuracy loss to the quantized representations.  
To preserve search accuracy, we adopt an over-fetching strategy that uses \(\textit{EF}>n_b\) to enlarge the candidate set returned by PIM. The value of \(\textit{EF}\) is selected empirically based on the study in Section~\ref{sec:eval:ablation} to trade off performance with accuracy.

\subsection{Asynchronous Pipelined Query Scheduling} \label{sec:host-pu}

Our compact index design makes the IVF cluster the fundamental unit of deployment: each cluster contains a self-contained search structure (i.e., neighbor IDs and RabitQ code/factor arrays) in PU-local memory, while full-precision vectors remain on the host for final reranking. 
This organization avoids fine-grained graph partitioning across PUs and turns query execution into a cluster-aware scheduling problem. Based on this organization, \sysname{} adopts a two-level scheduling strategy consisting of offline cluster placement and online mini-batch pipeline execution.

\subsubsection{\textbf{Scheduling Challenges on Commodity PIM}}


Before online execution, \sysname{} first places compact-index clusters onto PUs using a greedy load-balancing policy based on estimated or profiled access frequency~\cite{chen2024upanns,chen2024drim}. 
This step mitigates persistent hot spots caused by skewed cluster popularity and improves utilization across the PU array. Because the compact index substantially reduces the memory footprint of each cluster, the scheduler has more flexibility to balance load while respecting the PU-local memory budget. In this sense, the data-layout optimization of \textbf{O1} directly enables the scheduling flexibility required by \textbf{O2}.

\begin{figure}[t]
    \centering
    \begin{subfigure}{.24\textwidth}
    \centering
        \includegraphics[width=\linewidth]{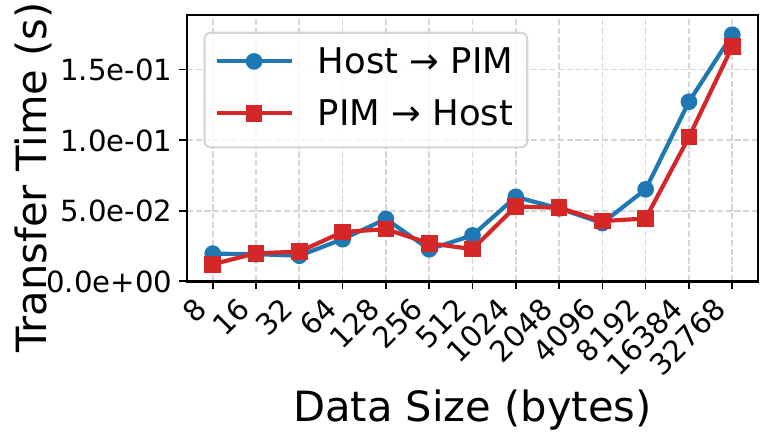}  
    \caption{UPMEM PIM}
    \label{fig:hp:upmem}
    \end{subfigure}
    \hfil
    \begin{subfigure}{.24\textwidth}
    \centering
        \includegraphics[width=\linewidth]{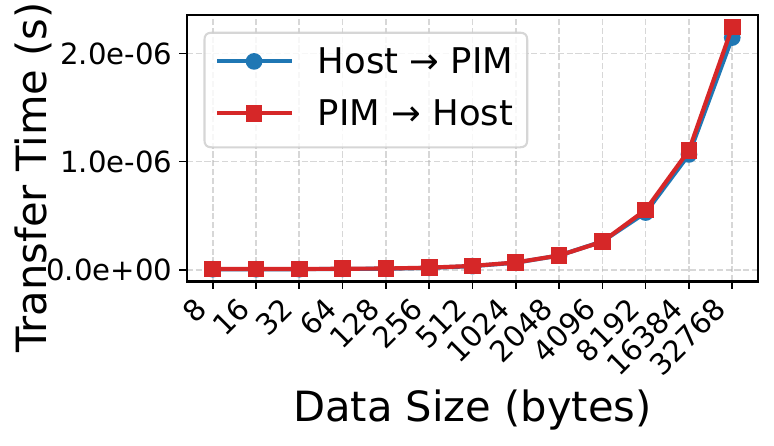}   
    \caption{PIM-HBM (Simulated)}
    \label{fig:hp:pim-hbm}
    \end{subfigure}
    \caption{Host-PIM communication overhead. We vary the size of data transferred from/to PIM and measure the latency.}
    \label{fig:hp}
\end{figure}

After cluster placement, the primary runtime bottleneck becomes host–PIM coordination, due to the significant gap between external and internal bandwidth on PIM. 
Figure~\ref{fig:hp} characterizes the communication cost between the host and PIM across different transfer sizes. The key observation is that transfer latency is not linear in practice: very small transfers under-utilize bandwidth, while large transfers incur much higher latency. 
Therefore, an effective scheduling policy must strike a balance between these two extremes.

\begin{figure}[t]
    \centering
    \includegraphics[width=.9\linewidth]{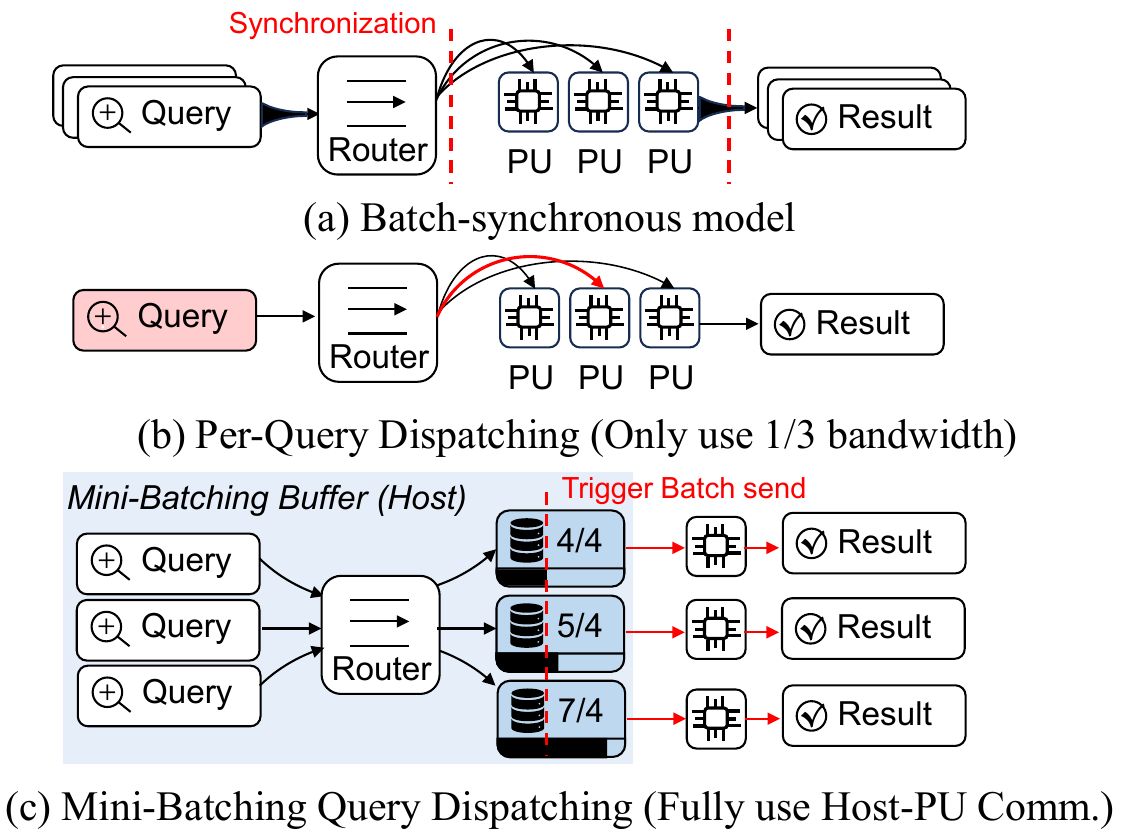}
    \caption{Comparison of different scheduling strategies.}
    \label{fig:mini-batching}
\end{figure}

Unfortunately, existing mainstream scheduling strategies fail to navigate this trade-off effectively.
The widely adopted \textbf{batch-synchronous} model~\cite{cui2025pimlex, cai2024pimpam, li2024pim,chen2024upanns,chen2024drim} aggregates a large batch of queries, dispatches them to PUs, and blocks until every PU finishes before collecting results (Figure~\ref{fig:mini-batching}(a)).
This introduces rigid global barriers: the host remains idle during in-PIM search, and the PUs later remain idle while the host performs post-processing and prepares the next batch. 
%
At the opposite extreme, \textbf{per-query dispatching} removes the global barrier by sending each query immediately (Figure~\ref{fig:mini-batching}(b)). However, this strategy fragments communication into transfers that are too small to efficiently utilize the host–PU bandwidth, and fails to exploit system-wide transfer parallelism when the host serializes many tiny requests. 

\subsubsection{\textbf{Dynamic Mini-Batch Pipeline Design}}

To balance communication efficiency and execution overlap, \sysname{} adopts \textbf{dynamic mini-batching}, as illustrated in Figure~\ref{fig:mini-batching}(c). 
The host maintains a input buffer per-PU, and each incoming query is appended to the buffers of the PUs that store its relevant clusters after cluster filtering. A mini-batch is dispatched when either the workload reaches a target \emph{threshold} (e.g., 4 in Figure~\ref{fig:mini-batching}(c)) or the oldest buffered query exceeds a waiting-time limit. This policy allows the system to aggregate enough work to amortize communication overhead under heavy load, while still being responsive when the arrival rate is low.

Based on this dispatch policy, \sysname{} organizes execution as an asynchronous pipeline spanning the host and PIM. Logically, the pipeline contains three components. First, the \emph{Query Dispatcher} on the host performs cluster filtering, fills per-PU buffers, and sends ready mini-batches to PIM. Second, the \emph{Search Engine} on each PU reads a mini-batch, traverses the compact PIM-resident graph index, and computes approximate distances using the multiplication-free kernel described in the next subsection. Third, the \emph{Post-processor} on the host continuously fetches returned candidate sets, retrieves the corresponding raw vectors, and performs exact reranking to produce the final top-$k$ results. This execution flow is consistent with the end-to-end workflow shown in Figure~\ref{fig:overview}.

\begin{figure}[t]
    \centering
    \includegraphics[width=\linewidth]{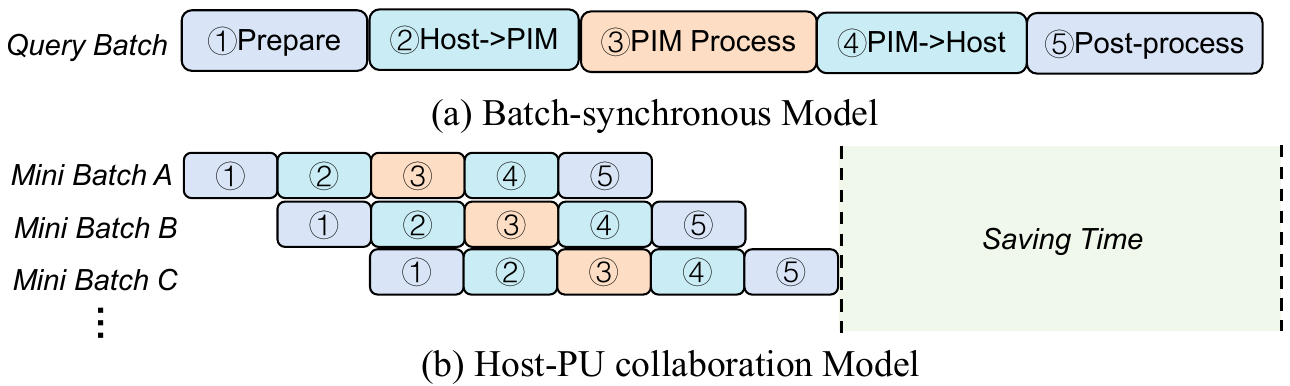}
    \caption{Comparison of Host-PU collaboration strategy.}
    \label{fig:pipeline}
\end{figure}

Although the logical execution consists of three components, the end-to-end processing is executed as five overlapped stages: \circled{1} host-side query preparation, \circled{2} host-to-PIM transfer, \circled{3} in-PIM query processing, \circled{4} PIM-to-host result return, and \circled{5} host-side reranking. Figure~\ref{fig:pipeline} highlights the key benefit of this design: unlike batch-synchronous execution, communication, in-PIM search, and host-side reranking proceed concurrently rather than serially. As a result, the host no longer waits for all PUs to finish before beginning post-processing, and the PUs no longer stall until the host completes reranking for an entire batch. This is particularly important for graph-based ANNS, where traversal varies significantly across queries.

To maintain stable execution, \sysname{} uses FIFO queues to decouple the host and PIM stages and applies lightweight flow control to bound the number of in-flight mini-batches. This avoids queue overflow without reintroducing coarse-grained synchronization barriers. 

The effectiveness of this design also depends on choosing a good mini-batch size. Let $T_{\mathrm{pre}}(N_B)$, $T_{\mathrm{proc}}(N_B)$, and $T_{\mathrm{post}}(N_B)$ denote the host-side dispatch time, in-PIM processing time, and host-side reranking time for a mini-batch of size $N_B$, respectively. Since these stages execute in parallel, the average processing time per query is determined by the slowest stage:
\begin{equation}
\small 
    T(N_B) = \frac{\max \left( T_{\mathrm{pre}}(N_B),\; T_{\mathrm{proc}}(N_B),\; T_{\mathrm{post}}(N_B) \right)}{N_B}
\end{equation}
Therefore, the optimal mini-batch size is chosen as {\small
\[
N^*=\underset{N_B}{\arg\min}\ T(N_B).
\]
}

Based on the real hardware limits observed in Figure~\ref{fig:hp}, we tune $N^*$ to the point where $T_\mathrm{proc} \approx \max\big(T_\mathrm{pre}, T_\mathrm{post}\big)$. This ensures that the data size stays within the fast communicating range (e.g., under 8\,KB), keeping the execution pipeline fully balanced and efficient.

\subsection{Multiplication-Free Distance Computation}\label{sec:multiply}

After the compact index and asynchronous pipeline are in place, the efficiency of the PU-side search engine is determined primarily by its inner-loop distance computation.

\sysname{} inherits the distance computation kernel from RabitQ~\cite{gao2024rabitq}, where 
the approximate distance between the query vector $\mathbf{q}$ and the candidate vector $\mathbf{o}$ is computed as:
\begin{align}\small
    d_{appro} &= \underbrace{\|\mathbf{o}\|^2 + 2 \cdot \left(\frac{\|\mathbf{o}\|}{\langle\mathbf{\bar{o}},\mathbf{o}\rangle}\right) \cdot \langle \mathbf{c}, \mathbf{\bar{o}} \rangle \cdot \sqrt{D}}_{\text{Query-independent term}} + \|\mathbf{q}\|^2 \nonumber \\
    &\quad - \underbrace{\left(\frac{\|\mathbf{o}\|}{\langle\mathbf{\bar{o}},\mathbf{o}\rangle}\right) }_{\text{Outer scaling}}{\color{red} \cdot }  \underbrace{\big(2 \cdot \langle \mathbf{\bar{o}}, \mathbf{q} \rangle - sumq\big) \cdot 2}_{\text{Reduced to additions via RabitQ}} \label{eq:second}
\end{align}
where $\mathbf{\bar{o}}$ is the vector reconstructed from the quantization code, $\mathbf{c}$ is the cluster centroid, and $sumq$ is the sum of query values. 

Conceptually, this formula consists of three logical components: (1) a query-independent term that can be precomputed, (2) an inner-product term heavily optimized into simple additions, and (3) an outer scaling operation that involves floating-point multiplication and division. 
Thus, although RabitQ successfully removes multiplication from the inner product itself, the remaining outer scalar is still poorly matched to commodity PIM cores and becomes a major source of latency inside the \emph{Search Engine}. 

To minimize runtime arithmetic, we first isolate all query-independent terms and precompute them as a single $\textit{RabitQFactor}$ during index construction. Since $\|\mathbf{q}\|^2$ is identical for all candidates within a query, it can be omitted without affecting relative ranking. In addition, all candidate vectors are normalized in advance, so $\|\mathbf{o}\|$=1. Under these conditions, the static portion of the approximate-distance expression can be absorbed into a per-node constant stored in the compact index.
The primary remaining computational bottleneck is therefore the outer scalar factor derived from $\langle\mathbf{\bar{o}},\mathbf{o}\rangle$. 

Because $\mathbf{\bar{o}}$ is reconstructed through a random orthogonal transformation, its norm is also 1, and the inner product $\langle\mathbf{\bar{o}},\mathbf{o}\rangle$ reduces to $cos(\theta)$, where $\theta$ is the angle between the original vector $\mathbf{o}$ and its reconstruction $\mathbf{\bar{o}}$. In the original RabitQ formulation, this term is node-specific, forcing the PU to repeatedly apply a distinct floating-point scaling factor for each candidate.
Our key observation is that this quantization-error term is sufficiently stable within an IVF cluster. Because vectors in the same cluster are encoded relative to a shared centroid, their quantization-error distribution is also empirically stable at the cluster level. This makes a cluster-wise approximation natural. Therefore, instead of maintaining a separate $cos(\theta)$ for each node, \sysname{} replaces it with a cluster-wide constant $\alpha$, yielding the simplified approximate-distance formulation:
\begin{align}\small
    d_{appro} = \textit{RabitQFactor} - \frac{1}{\alpha} \cdot (RabitQ\_result) \label{eq:newdistance}
\end{align}
where $RabitQ\_result$ denotes the lookup-based term already computed using additions.

To fully eliminate multiplication from the online search path, we further approximate the inverse factor $1/\alpha$ using only bit shifts and additions. Empirical analysis indicates that for typical feature dimensions (e.g., $10^2$ to $10^6$), $cos(\theta)$ concentrates around 0.8~\cite{gao2024rabitq}. Setting $\alpha $= 0.8 gives $1/\alpha$ = 1.25. This value has a convenient binary representation, $1.01_2$, which translates directly into efficient bitwise operations: $ x \cdot 1.25 \approx x + (x >> 2)$.
When a specific dataset or cluster deviates from this default, $\alpha$ is calibrated during index construction to the nearest hardware-friendly binary-shift equivalent. As a result, the expensive floating-point scaling is removed from online search and absorbed into lightweight offline preprocessing.


\begin{figure}[t]
    \centering
    \includegraphics[width=0.95\linewidth]{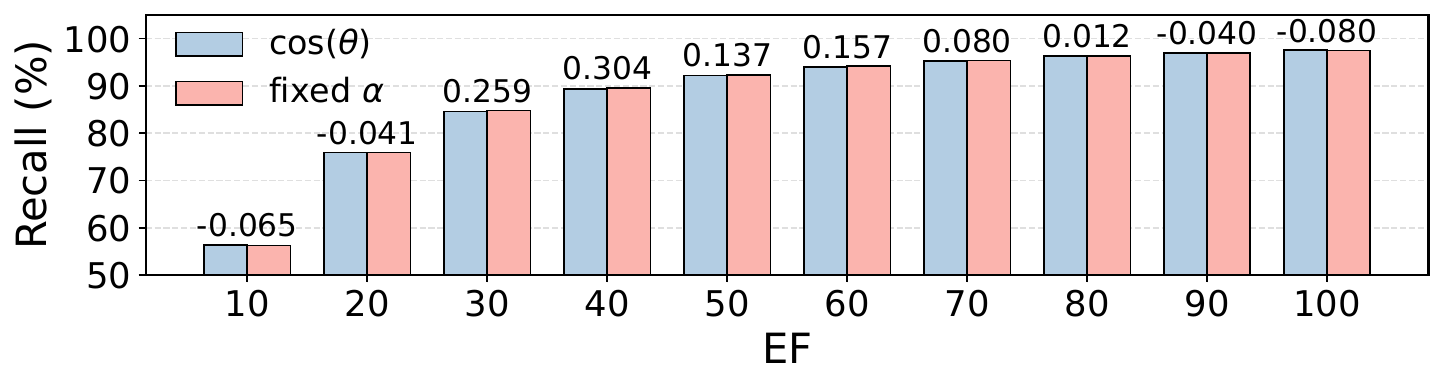}
    \caption{Recall of \sysname{} using node-specific $cos(\theta)$ or fixed $\alpha$ on SIFT.}
    \label{fig::mf:rcall}
\end{figure}

Finally, we validate that this simplification does not materially degrade search quality. Figure~\ref{fig::mf:rcall} compares the original node-specific $cos(\theta)$ formulation against the fixed-$\alpha$ version across different $\textit{EF}$ settings on the SIFT dataset. The results show that using a fixed $\alpha$ = 0.8 incurs negligible accuracy loss, with a maximum recall drop below 0.08\%. Therefore, PIMCQG achieves a multiplication-free in-PIM distance kernel while preserving the ranking quality for high-recall graph traversal.

\section{Experimental Evaluation}


\begin{figure*}[t]
    \centering
    \includegraphics[width=\linewidth]{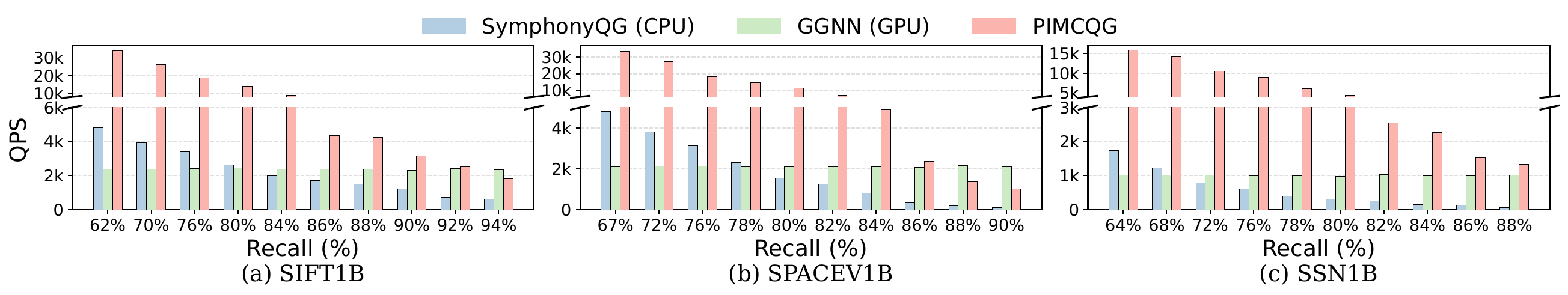}
    \caption{QPS vs. recall@10 of compared baselines. Each point is obtained by varying the search-cluster count and $\textit{EF}$.}
    \label{fig:graph-comp}
\end{figure*}

\begin{figure*}[t]
    \centering
    \includegraphics[width=\linewidth]{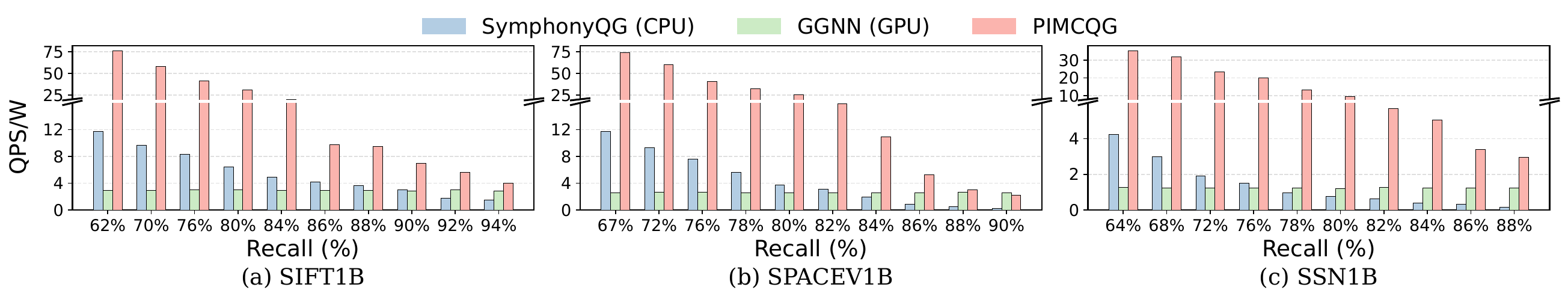}
    \caption{Energy efficiency (QPS/W) vs. recall@10 for the compared baselines across three datasets. }
    \label{fig:graph-comp-qpsw}
\end{figure*}

We evaluate \sysname{} on a real-world commodity PIM platform to answer three key questions: (1) Can \sysname{} deliver high-throughput graph-based ANNS at high recall, (2) how do its algorithm–hardware co-design choices contribute to performance and energy efficiency, and (3) does the design scale across datasets and system configurations.
\subsection{Experimental Setup}

\subsubsection{\textbf{Hardware Setup}}
We conduct evaluations on three types of hardware platforms, including PIM, CPU and GPU.
The \textbf{PIM platform} contains a real-world UPMEM PIM server~\cite{devaux2019true} equipped with a dual-socket Intel Xeon Silver 4110 host processor (2.10GHz), 256GB DDR4 host memory, and 20 PIM modules providing up to 2,560 DPUs operating at 350 MHz. The total system power is approximately 450W, including $\sim$170W for the host CPU and $\sim$14W per activated PIM module.
For future-looking scalability analysis, we additionally use the vendor-provided open-source simulators for Samsung PIM-HBM~\cite{pim-hbm} and SK Hynix AiM~\cite{aim}, since these platforms are not yet publicly accessible as commercial systems. 
The \textbf{CPU platform} contains a dual-socket Intel Xeon Gold 6330 platform (2.0GHz, 112 threads) with 512GB of host memory and a total system power of $\sim$410W. The \textbf{GPU platform} attaches an NVIDIA A100-SMX4 (80GB) to the CPU platform, resulting in an estimated total system power of $\sim$810W. For multi-GPU comparison, we expand the system to include up to eight A100 GPUs with the same configurations.

\subsubsection{\textbf{Datasets}}
We experiment on three industry-standard billion-scale benchmarks: SPACEV1B ($D$=100)~\cite{spacev1b}, SIFT1B ($D$=128)~\cite{sift1b}, and SimSearchNet++ (SSN1B, $D$=256)~\cite{ssn1b}. 
Their uncompressed raw-vector footprints are 95GB, 123GB, and 239GB, respectively. For each dataset, we use the default public query set and compute the ground-truth nearest neighbors using exact brute-force search.

\subsubsection{\textbf{Baselines}}
We compare \sysname{} against four representative baselines spanning CPU, GPU, and PIM platforms.   
\begin{itemize}[leftmargin=*]
    \item \textbf{SymphonyQG}~\cite{gou2025symphonyqg} serves as the primary CPU baseline and the SOTA graph-based ANNS method. Because the full SymphonyQG index exceeds the memory capacity of the CPU platform, we apply the same IVF partitioning used by PIMCQG and load only the query-relevant clusters into host memory; disk I/O is excluded from all reported timings to isolate search performance.
    \item \textbf{UpANNS}~\cite{chen2024upanns} and \textbf{PIMANN}~\cite{wu2025turbocharge} serve as the two PIM baselines, representing batch-synchronous and per-query scheduling strategies, respectively, on UPMEM. 
    \item \textbf{GGNN}~\cite{groh2023ggnn}, a popular GPU-based acceleration for billion-scale graph-based ANNS, is adopted for comparison since SymphonyQG lacks GPU support. We select GGNN over newer GPU-based systems (e.g., CAGRA~\cite{ootomo2024cagra}, PathWeaver~\cite{kim2025pathweaver}), which target million-scale datasets and do not scale to billion-scale workloads. 
    GGNN utilizes a $k$-NN graph index, similar to SymphonyQG and \sysname. For fairness on billion-scale datasets, GGNN is configured with 16 data shards in our experiments.
\end{itemize}

\subsubsection{\textbf{Metrics}} 
We report four primary metrics: throughput (QPS), recall@10, index size (GB), and energy efficiency (QPS/W). Unless otherwise specified, our default setting uses SIFT1B, builds an IVF partitioning with 8,192 clusters (limited by the 64 MB per-DPU memory budget), probes 8 clusters per query, constructs the graph index with node degree 32, and sets the over-fetched candidate size $\textit{EF}$ to 40. 


\subsection{End-to-End Performance Across Hardware}

We first evaluate whether \sysname{} improves the end-to-end throughput and energy efficiency of graph-based ANNS relative to representative CPU and GPU baselines. Specifically, we compare \sysname{} against SymphonyQG on CPU and GGNN on GPU, and vary the search-cluster count and $\textit{EF}$ to cover a broad range of recall targets. Figures~\ref{fig:graph-comp} and \ref{fig:graph-comp-qpsw} report the resulting throughput and energy-efficiency trade-offs.

Across most recall targets, \sysname{} achieves the highest throughput among all three platforms. Compared with SymphonyQG, \sysname{} delivers up to 7.1×, 7.4×, and 20× higher QPS on SIFT1B, SPACEV1B, and SSN1B, respectively. Compared with single-GPU GGNN, \sysname{} achieves up to 17.1×, 16.7×, and 15.8× higher throughput on the same datasets within the practically relevant recall range of approximately 0.60–0.84. At very high recall targets, GGNN becomes more competitive, especially on SPACEV1B. This behavior is expected because the single A100 GPU cannot fully hold the full graph index, forcing CPU–GPU data swapping that becomes a persistent overhead regardless of search strictness. Even so, \sysname{} maintains a clear advantage throughout the main operating region used in practice.

\sysname{} also provides substantially higher energy efficiency than both baselines. On SIFT1B, it achieves 4–76 QPS/W, compared with 1.5–11 QPS/W for SymphonyQG and about 2.5 QPS/W for GGNN, corresponding to improvements of up to 6.5× over CPU and 30.8× over GPU. Similar trends hold on SPACEV1B and SSN1B, confirming that the throughput gains of \sysname{} are not achieved through disproportionate power consumption, but instead through a more energy-proportional execution model.

\begin{figure}[t]
    \centering
    \includegraphics[width=\linewidth]{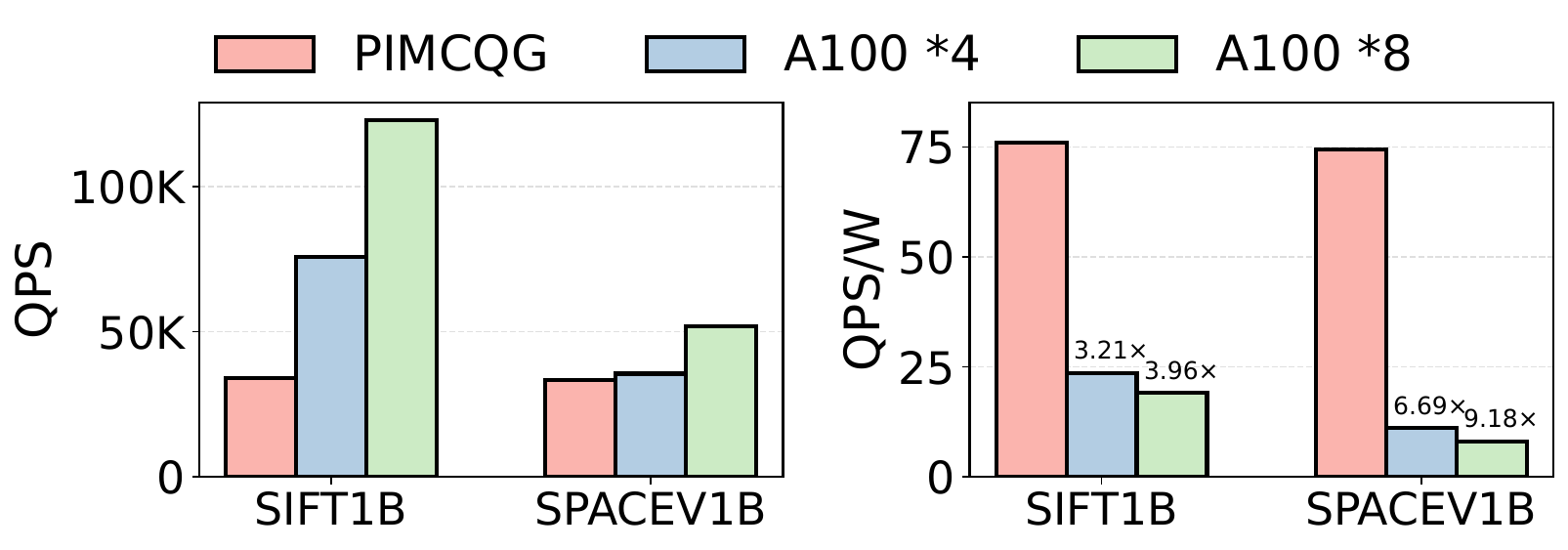}
    \caption{Comparing \sysname{} with GGNN on 4 and 8 GPUs.}
    \label{fig:ab:multigpu}
\end{figure}

To isolate the effect of GPU memory capacity, we additionally evaluate GGNN on 4× and 8× A100 configurations in which the full index fits in GPU memory. As shown in Figure~\ref{fig:ab:multigpu}, multi-GPU GGNN significantly increases raw throughput and can surpass \sysname{} in absolute QPS. However, this comes at a much higher system power cost. As a result, \sysname{} still retains a 3.2×–9.1× advantage in QPS/W over the 4× and 8× A100 configurations on SIFT1B and SPACEV1B. These results show that while aggressively scaled GPU systems can deliver high peak throughput, \sysname{} offers a substantially more energy-efficient solution for billion-scale graph-based ANNS.

\subsection{Comparison with Existing PIM-based ANNS Systems}
\begin{figure}[t]
    \centering
    \includegraphics[width=\linewidth]{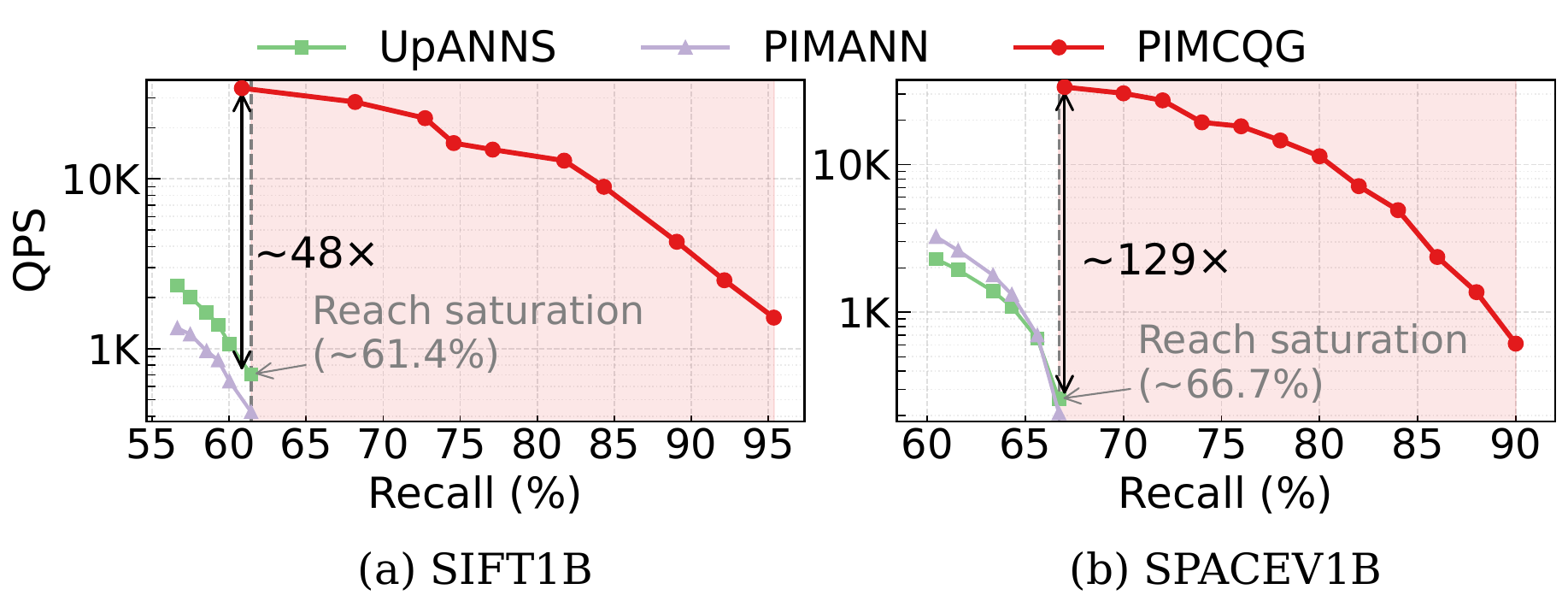}
    \caption{QPS vs. recall@10 for PIMCQG and prior PIM-based ANNS systems.}
    \label{fig:pim-comp}
\end{figure}

We next compare \sysname{} with prior PIM-based ANNS systems to assess whether the proposed co-design closes the capability gap between PIM and \emph{high-recall} graph-based search. We use UpANNS and PIMANN as representative baselines, both of which implement IVFPQ-based ANNS~\cite{chen2024upanns,chen2024drim} on the UPMEM platform. Since their original studies used a relaxed recall definition, we re-evaluate both methods using the standard recall@10 metric to ensure a fair comparison. Figure~\ref{fig:pim-comp} reports the resulting recall-throughput curves.

Under this standardized evaluation protocol, \sysname{} shows a much stronger recall-throughput trade-off than both prior PIM solutions. A key observation is that UpANNS and PIMANN reach a clear capability ceiling at relatively low recall levels (e.g., 61.4\% on SIFT1B and 66.7\% on SPACEV1B), after which throughput drops sharply. In contrast, \sysname{} maintains high throughput well beyond these saturation points. 

When compared at recall levels near the saturation boundaries of the prior PIM baselines, \sysname{} achieves up to 48× higher throughput on SIFT1B and 129× higher throughput on SPACEV1B. These results show that the contribution of \sysname{} is not merely incremental acceleration over previous PIM designs; rather, it enables a qualitatively different operating regime by supporting the higher recall levels expected from modern graph-based ANNS workloads.



\subsection{Component-wise Analysis and Ablation}\label{sec:eval:ablation}

To understand where the gains of \sysname{} come from, we next evaluate the individual effects of its three co-designed optimizations and analyze the remaining system bottlenecks. 

\subsubsection{Compact Index Footprint}

\begin{table}[t]
    \centering
    \caption{Index footprint of \sysname{} and SymphonyQG.}
    \label{tab:index_size}
    \small 
    \setlength{\tabcolsep}{4pt} 
    \begin{tabular}{lccc} 
        \toprule
                        & SIFT1B & SPACEV1B & SSN1B \\ 
        \midrule
        SymphonyQG      & 1423 GB & 1327 GB       & 2385 GB             \\
        \sysname{}          & 138 GB  & 138 GB       & 164 GB             \\
        Reduction Ratio & 10.31x  & 9.62x       & 14.54x             \\ 
        \bottomrule
    \end{tabular}
\end{table}

We first examine the memory footprint of the compact index introduced in Section~\ref{sec:index}. Table~\ref{tab:index_size} compares the index size of \sysname{} with SymphonyQG across the three billion-scale datasets. By removing raw vectors from PIM and replacing per-edge metadata with a compact cluster-aware structure, \sysname{} reduces index size by 10×–14× (e.g., from 1423GB to 138GB on SIFT1B). This reduction is critical for making billion-scale graph-based ANNS feasible on commodity PIM and directly validates the effectiveness of the compact-index design.

\subsubsection{Pipeline Bottleneck Analysis}

\begin{figure}[t]
    \centering
    \includegraphics[width=0.95\linewidth]{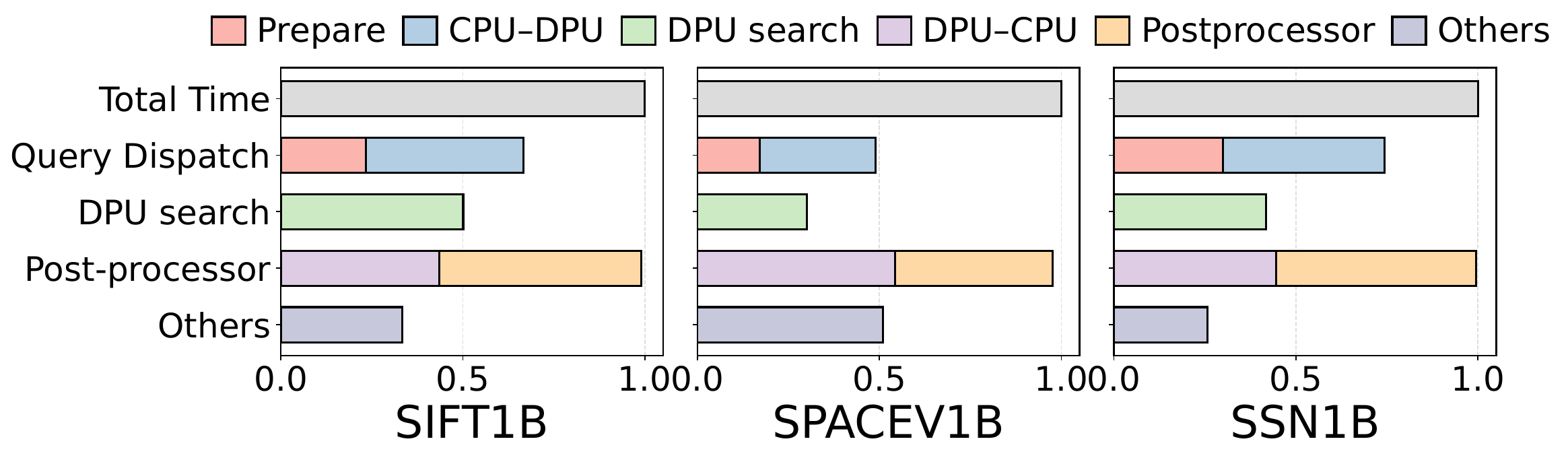}
    \caption{Performance breakdown. The overall execution time is dictated by the slowest pipeline stage.}
    \label{fig:breakdown}
\end{figure}

We then analyze the runtime breakdown of \sysname{} using Figure~\ref{fig:breakdown}. 
Since the asynchronous pipelined execution overlaps query dispatch, DPU search, and post-processing, the end-to-end latency is determined by the slowest pipeline stage. 
Across all datasets, the actual DPU search contributes only a relatively small fraction ($\leq$50\%), 
while the post-processing stage, including DPU-to-CPU result transfer and host-side exact distance recomputation, dominates the total execution time. Note that this overhead is attributed to two inherent factors: offloading raw 
vectors to the host to reduce PIM memory footprint, and the limited host-PIM bandwidth of current UPMEM hardware.
It also suggests that \sysname{} can benefit substantially from future PIM systems with higher host-PIM bandwidth.

\begin{figure}[t]
    \centering
    \includegraphics[width=0.95\linewidth]{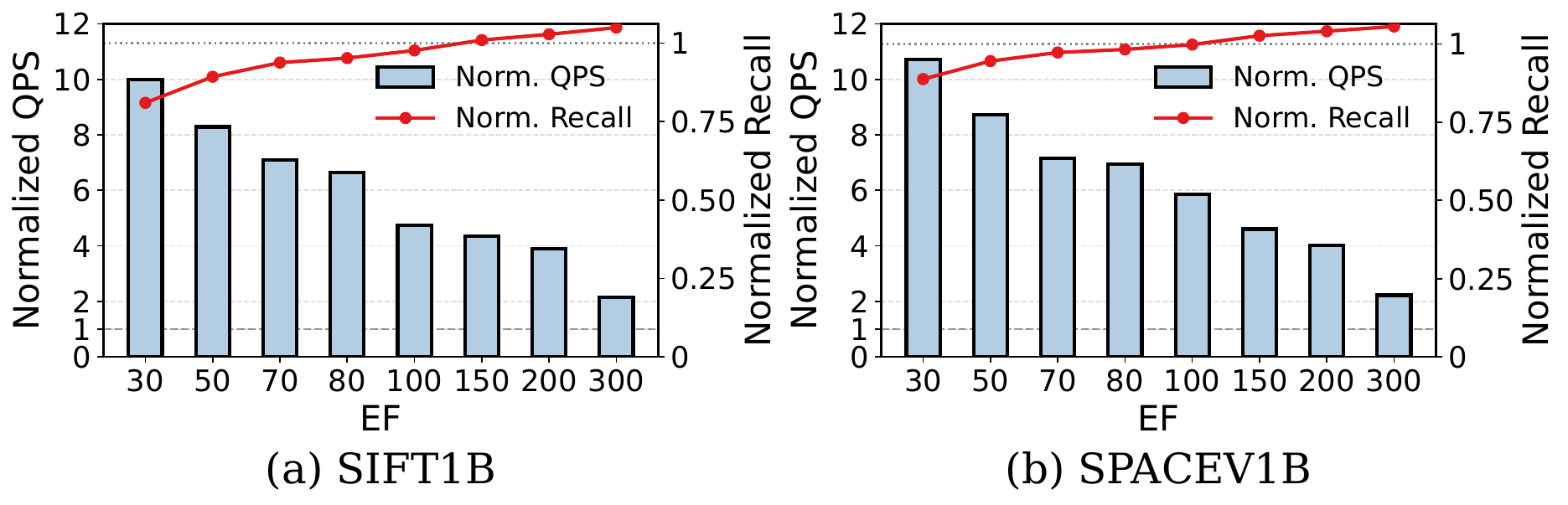}
    \caption{The impact of overfetching-reranking strategy}
    \label{fig:ab:overfetch}
\end{figure}

\subsubsection{Overfetching-Reranking} 

\sysname{} relies on a host-side reranking stage over an overfetched candidate set (size of $\textit{EF}$) to preserve high accuracy. To study the impact of this strategy, we vary the value of $\textit{EF}$ and normalize the QPS and recall results of \sysname{} against the results of SymphonyQG with the candidate set size $n_b$=30.

Figure~\ref{fig:ab:overfetch} clearly shows the trade-off between throughput and accuracy. Without overfetching ($\textit{EF}=n_b$), \sysname{} achieves very high throughput (10×-10.4× that of SymphonyQG), but achieves only 81\%–89\% of the baseline recall. 
When increasing the overfetch size (to 150 for SIFT1B and 100 for SPACEV1B), \sysname{} achieves the same recall level as SymphonyQG, while still preserving 4×-6x higher QPS. This result confirms that overfetching-reranking is an effective mechanism for preserving accuracy without sacrificing the throughput advantage of the compact PIM index.


\begin{figure}[t]
    \begin{minipage}{.23\textwidth}
    \centering
        \includegraphics[width=\linewidth]{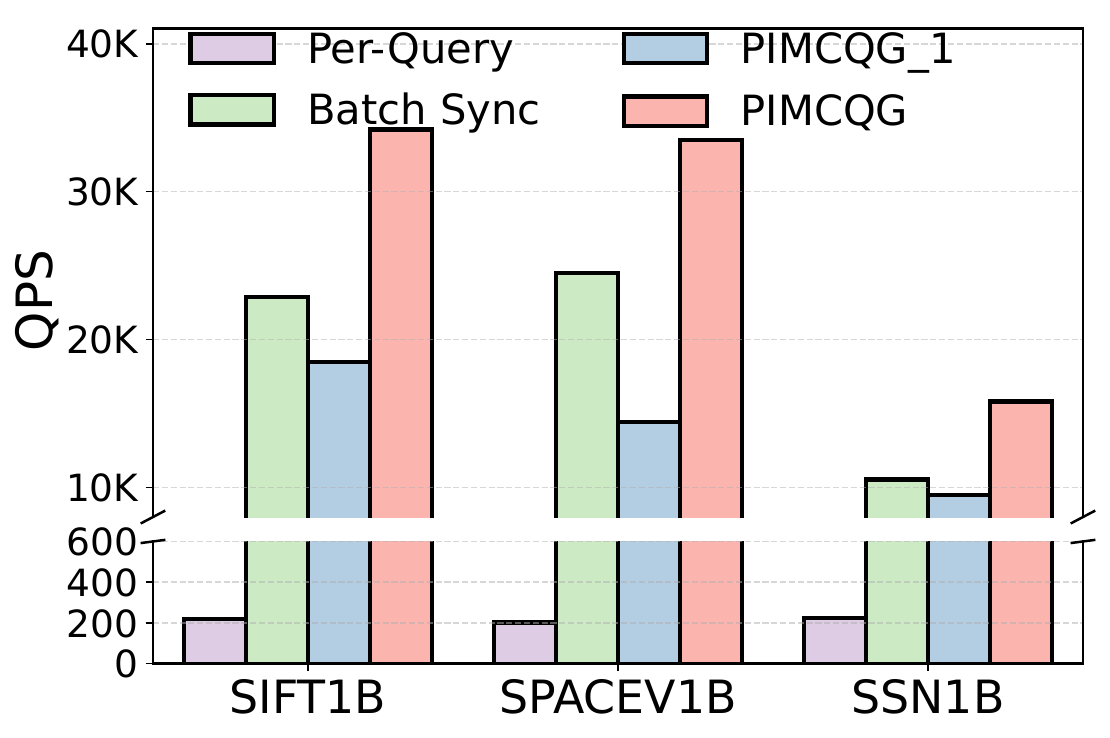}
    \caption{Throughput comparison of different scheduling strategies}\label{fig:ab:async}
    \end{minipage}
    \hfill
    \begin{minipage}{.23\textwidth}
    \centering
        \includegraphics[width=\linewidth]{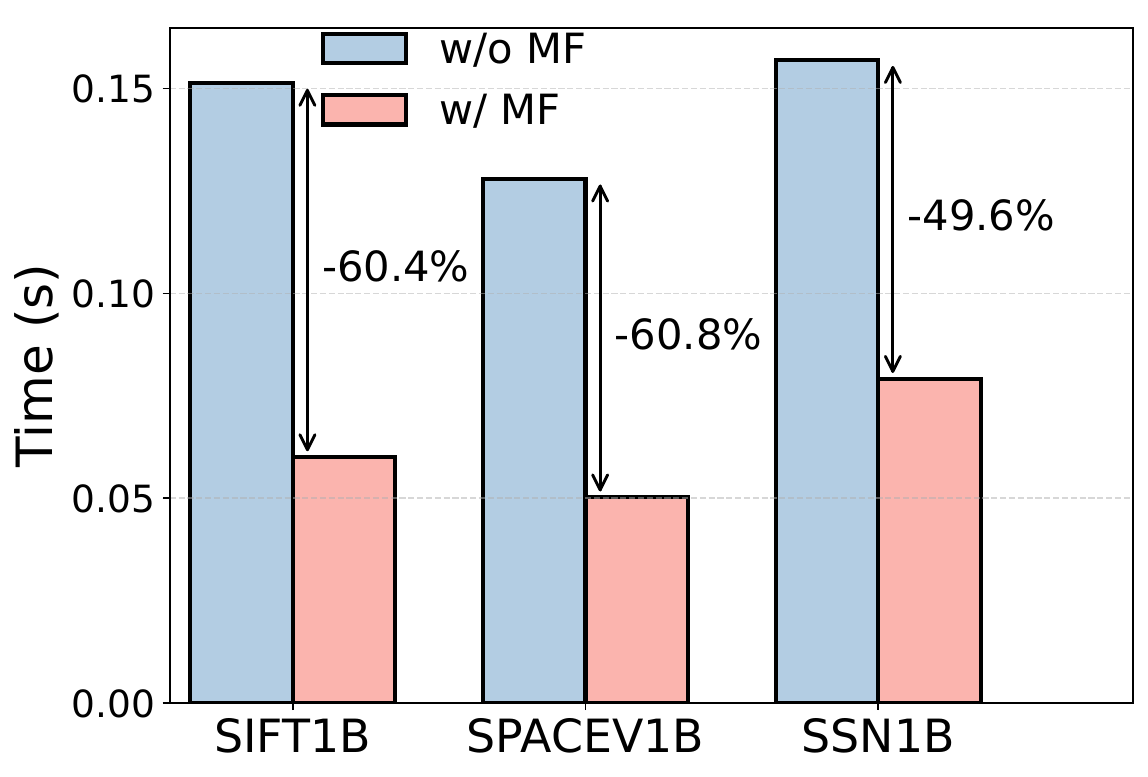}
    \caption{DPU search time w/ and w/o multiplication-free distance computation}
    \label{fig:ab:mf}
    \end{minipage}
\end{figure}

\subsubsection{Asynchronous Pipelined Scheduling} 

We next isolate the contribution of the scheduling design in Section~\ref{sec:host-pu} by comparing \sysname{} against three alternatives: per-query dispatching, batch-synchronous scheduling, and PIMCQG\_1, a pipeline variant with mini-batch size fixed to one.

Figure~\ref{fig:ab:async} shows that \sysname{} outperforms naive per-query dispatching by 70×–155× across all three datasets, demonstrating that fine-grained dispatch fails to utilize host–DPU communication parallelism. 
Compared with batch-synchronous execution, the full asynchronous pipeline yields a $\sim$1.5× throughput improvement by overlapping communication and computation across query batches. Finally, relative to PIMCQG\_1, where the mini-batch size is fixed to 1, using an appropriately sized mini-batch provides a further 1.7×–2.4× improvement, 
confirming that dispatch granularity is critical for approaching peak effective bandwidth.



\subsubsection{Multiplication-free Distance Computation} 

Finally, we evaluate the effect of our multiplication-free distance kernel. Figure~\ref{fig:ab:mf} compares the DPU search phase of \sysname{} with and without the shift-add reformulation enabled. Across the three datasets, the optimized kernel reduces DPU search time by 49.6\%–60.8\%. This confirms that even after data movement and scheduling are optimized, arithmetic simplification remains essential for commodity PIM, and that removing floating-point multiplication from the PU-side critical path is a major contributor to the overall performance of \sysname{}.



\subsection{Scalability}
\begin{figure}[t]
    \begin{minipage}{.24\textwidth}
    \centering
        \includegraphics[width=0.95\linewidth]{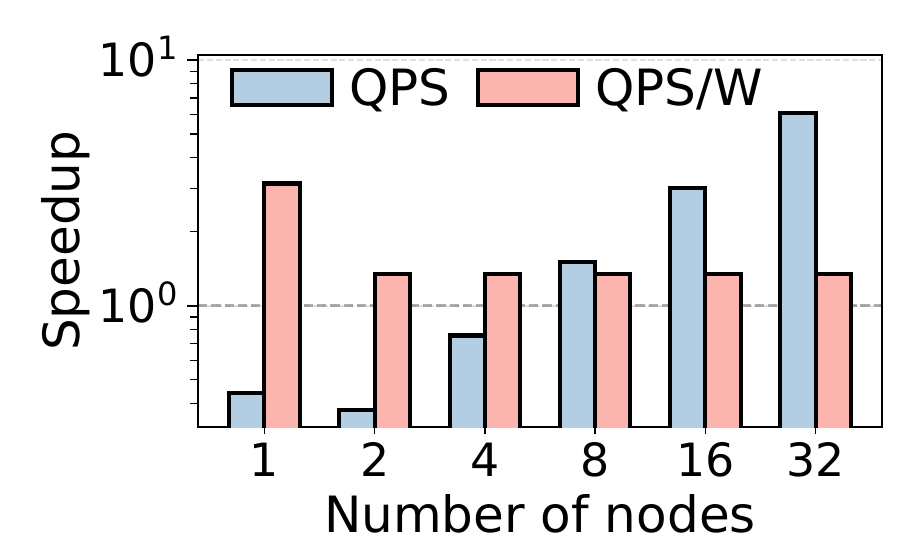}
    \caption{Multi-node scalability of \sysname{} on SIFT1B.}
    \label{fig:scale:a}
    \end{minipage}
    \begin{minipage}{.24\textwidth}
    \centering
        \includegraphics[width=0.95\linewidth]{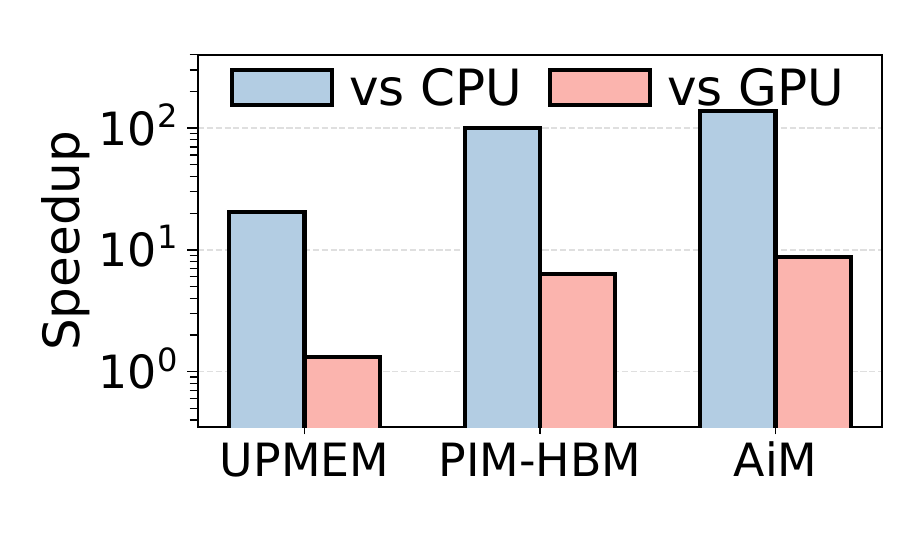}
    \caption{Speedup of \sysname{} on different PIM architectures.}
    \label{fig:scale:b}
    \end{minipage}
\end{figure}

We finally evaluate the scalability of \sysname{} in two dimensions: scale-out across multiple nodes and portability to emerging PIM hardware. 

\subsubsection{Multi-node Scalability}

To evaluate distributed scaling, we simulate a multi-node deployment of \sysname{} using a 400 Gbps InfiniBand network model in which communication cost scales with data transfer size. All speedups are normalized to GGNN on 4× A100 GPU for SIFT1B.
Figure~\ref{fig:scale:a} shows that scaling from one node to two nodes initially causes a performance drop due to inter-node communication overhead.
However, as the system scales from 2 to 32 nodes, \sysname{} exhibits near-linear speedup. At that point, the large amount of query-level parallelism dominates the inter-node overhead, indicating that \sysname{} can effectively exploit scale-out execution at the cluster level.



\subsubsection{PIM-architecture Scalability}

We also project \sysname{} onto two emerging PIM architectures, Samsung PIM-HBM and SK Hynix AiM, both of which provide substantially higher operating frequency and host–PIM bandwidth than current UPMEM systems. Using the vendor-provided simulators, we model the search time of \sysname{} with a GEMV kernel that matches the computational complexity of the optimized distance computation and scale it by the empirically observed average number of graph-search hops per query. 
Figure~\ref{fig:scale:b} shows that on these future platforms, \sysname{} is projected to achieve 100×–137× speedup over the CPU baseline and 6.3×–8.7× speedup over the 4× A100 GPU baseline on SIFT1B. These projections suggest that the algorithmic structure of \sysname{} is well aligned with future high-bandwidth PIM systems and can benefit directly as the underlying hardware matures.


\section{Related Work}
\label{sec:related_work}

We categorize related works into two main areas: hardware acceleration for ANNS and the broader adoption of PIM. 

\textbf{Hardware Acceleration for ANNS.} 
Conventional CPUs and GPUs remain common choices for ANNS acceleration, but their performance is fundamentally constrained by memory bottlenecks. 
CPU-based methods~\cite{gou2025symphonyqg,gao2024rabitq,wang2025accelerating,chen2023finger} accelerate distance computation via SIMD but remain memory bandwidth-bound. 
Conversely, GPUs offer superior memory bandwidth~\cite{kim2025pathweaver,khan2024bang,ootomo2024cagra}. However, GPU acceleration is severely limited by small global memory capacities. 

To bypass these limits, various heterogeneous and specialized systems have been proposed. Heterogeneous storage systems~\cite{li2025scalable,tian2024fusionanns,jayaram2019diskann,chen2021spann} utilize multi-tier memory hierarchies, but they introduce complex data management and I/O synchronization overheads. Similarly,
specialized hardware solutions, including FPGA~\cite{zeng2023df, liang2022vstore, song2025efficient, jiang2023co}, SmartSSD~\cite{kim2022accelerating, tian2024scalable}, CXL-based memory pools~\cite{jang2023cxl, ko2025cosmos}, and NAND flash~\cite{wang2024ndsearch,xu2026proxima}, have been developed to push performance limits. These approaches are orthogonal to our PIM-centric design, which leverages commodity hardware without custom fabrication.

\textbf{PIM-Based ANNS and General Applications.}
PIM has emerged as a proven paradigm for overcoming memory bottlenecks, demonstrating significant success across data-intensive domains like databases~\cite{kong2025pimbeam, cai2024pimpam, cui2025pimlex}, large language models~\cite{ortega2024pim, zhao2025nl,gu2025pim,liu2025heterrag}, and recommendation systems~\cite{liu2023accelerating, lee2024cost, ke2021near, asgari2021fafnir}. However, in the context of ANNS, existing PIM accelerators~\cite{chen2024upanns, wu2025turbocharge, chen2024drim} target cluster-based algorithms (e.g., IVF-PQ), which provide lower search quality and throughput compared to graph-based methods. Our work bridges this gap by explicitly mapping the high-performance graph-based ANNS algorithm to real-world PIM hardware. 
\section{Conclusion}
We presented \sysname{}, an algorithm--hardware co-design framework that enables high-recall graph-based ANNS on commodity PIM by jointly redesigning data layout, scheduling, and distance computation.
The compact index reduces PIM-resident footprint by up to $14.5\times$, making billion-scale graphs feasible on distributed PU memories. 
The asynchronous pipeline overlaps host--PIM execution while fully utilizing host--PIM bandwidth.
The multiplication-free kernel removes all floating-point multiplications, cutting DPU search time by up to 60.8\%.
Our evaluation shows that \sysname{} delivers up to $20\times$/$17.1\times$ throughput over CPU/GPU baselines, $129\times$ over prior PIM systems at high recall, and maintains strong scalability across multi-node configurations and emerging PIM architectures.
Our bottleneck analysis shows that host--PIM bandwidth remains the primary constraint, suggesting that \sysname{} will benefit directly from emerging higher-bandwidth PIM architectures.

\section*{Acknowledgements}

This work is supported by the Hong Kong RGC GRF grant (No. 12201026), Guangdong and Hong Kong Universities ``1+1+1'' Joint Research Collaboration Scheme (Project No. 2025A0505000012) and a Tencent gift fund. 
We employed ChatGPT to refine the language of the entire
manuscript. All generative outputs were manually reviewed
and edited for technical accuracy.


\bibliographystyle{IEEEtranS}
\bibliography{refs}

\end{document}